\documentstyle[11pt,epsf]{article}
\input epsf.tex
 1
 1
 1

\def\lsim{\mathrel{\raise.3ex\hbox{$<$\kern-.75em\lower1ex\hbox{$\sim$}}}}
\def\gsim{\mathrel{\raise.3ex\hbox{$>$\kern-.75em\lower1ex\hbox{$\sim$}}}}
\begin{document}
\noindent
\thispagestyle{empty}
\renewcommand{\thefootnote}{\fnsymbol{footnote}}
\begin{flushright}
{\bf McGill/97-15}\\
{\bf UCSD/PTH 97-15}\\
{\bf hep-ph/9707337}\\
{\bf July 1997}\\
\end{flushright}
\vspace{.5cm}
\begin{center}
  \begin{Large}\bf
The Single Photon Annihilation Contributions to the \\[2mm] 
Positronium Hyperfine Splitting to Order $m_e\alpha^6$
  \end{Large}
  \vspace{1.5cm}

\begin{large}
 A.H.~Hoang$^{a}$,
 P.~Labelle$^{b}$ and
 S.M.~Zebarjad$^{b}$
\end{large}

\vspace{1.5cm}
\begin{it}
${}^a$ Department of Physics, University of California, San Diego,\\
   La Jolla, CA 92093-0319, USA\\[.5cm]
${}^b$  Department of Physics, McGill University,\\
   Montr\'eal, Qu\'ebec, Canada H3A 2T8
\end{it}

  \vspace{2.5cm}
  {\bf Abstract}\\
\vspace{0.3cm}

\noindent
\begin{minipage}{14.0cm}
\begin{small}
The single photon annihilation contributions for the positronium
ground state hyperfine splitting are calculated
analytically  to order
$m_e\alpha^6$ using NRQED. Based on intuitive physical arguments the
same result can also be determined by a trivial calculation using
results from existing literature. Our result completes the hyperfine
splitting calculation to order $m_e\alpha^6$.
We compare the theoretical prediction with the most recent experimental 
measurement.
\end{small}
\end{minipage}
\end{center}
\setcounter{footnote}{0}
\renewcommand{\thefootnote}{\arabic{footnote}}
\vspace{1.2cm}
%
%
%
\newpage
\noindent
Positronium, a two-body bound state consisting of an electron and
a positron, belongs to the first systems studied within the
relativistic quantum theory developed by Dirac. The existence of
positronium was predicted in 1934~\cite{Mohorovicic1} and experimentally
verified at the beginning of the 1950s~\cite{Deutsch1}. For the ground
state hyperfine splitting (hfs), the mass difference between the
$1{}^3\!S_1$ and $1{}^1\!S_0$ state, steadily improved experimental
measurements have meanwhile reached a precision of 3.6 ppm~\cite{Ritter1}
which makes the calculation of all ${\cal{O}}(\alpha^2)$ corrections to
the leading and next-to-leading order expression 
mandatory. So far only the order $\alpha^2\ln\alpha^{-1}$ corrections
have been fully determined~\cite{CaswellBodwin}.
Including also the known ${\cal{O}}(\alpha^3 \ln^2\alpha^{-1})$
corrections~\cite{Karshenboim1,thesis}
the theoretical expression for the
hfs reads\footnote{
We use natural units, in which $\hbar = c = 1$.
}  
\begin{equation}
W \, = \, 
m_e\,\alpha^4\,\bigg[\,
\frac{7}{12} - \frac{\alpha}{\pi}\,\bigg(\,
\frac{8}{9}+\frac{1}{2}\,\ln 2
\,\bigg)
+ \alpha^2\,\bigg(\,
\frac{5}{24}\,\ln\alpha^{-1} + K
\,\bigg) 
-\frac{7}{8 \pi}\,\alpha^3\,\ln^2\alpha^{-1}
\,\bigg]
\,.
\label{LOandNLO}
\end{equation}
We have calculated the single photon annihilation ($1-\gamma$ ann)
contributions to the
constant $K$. All other contributions to $K$ coming from the
non-annihilation, and the two and three photon annihilation processes
have been calculated before (see Table~\ref{tab1}). Although
our result completes the calculations to order $m_e \alpha^6$ the
theoretical situation remains unresolved due to a discrepancy in two
older calculations for some of the non-annihilation contributions to order
$m_e \alpha^6$.
\par
In this letter we report on two methods to determine 
$W^{\mbox{\tiny 1-$\gamma$ ann}}_{\mbox{\tiny NNLO}}$. The first one is
systematic using NRQED~\cite{Caswell1}, which is based on the concept of
effective field theories and the separation of effects from
non-relativistic  and relativistic momenta, and the second one
relies on physical intuition using two results from already
existing literature. We also summarize the status of the theoretical
calculation to the hfs in view of the most recent available
experimental data. We would like to note that the presentation of our
NRQED calculation is only meant to illustrate the basic steps of our
calculation. A more detailed and explicit work on the NRQED method is
in preparation. Our second method, on the other hand, is almost
trivial and represents a true ``back of the envelope'' calculation.
\par
\begin{figure}
\centerline{\epsfxsize 4.5 truein
\epsfbox{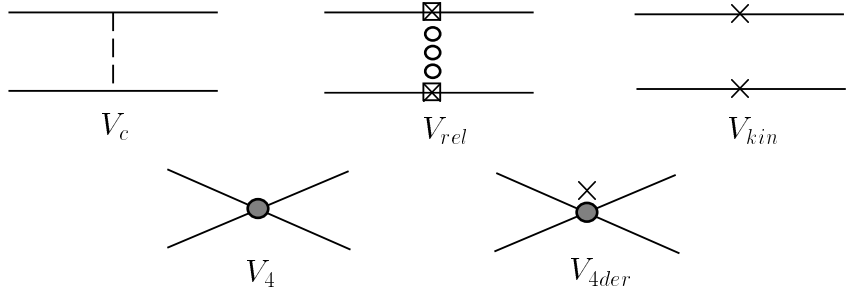}}
\caption{ 
Interaction potentials contributing to $W^{\mbox{\tiny 1-$\gamma$
ann}}_{\mbox{\tiny NNLO}}$. $V_{\mbox{\tiny kin}}$ denotes the
relativistic kinetic energy correction coming from the 
${\mbox{\boldmath $D$}}^4/8m_e^3$ terms in the NRQED Lagrangian.
}
\label{fig1}
\end{figure}
\begin{figure}
\centerline{\epsfxsize 6.2 truein
\epsfbox{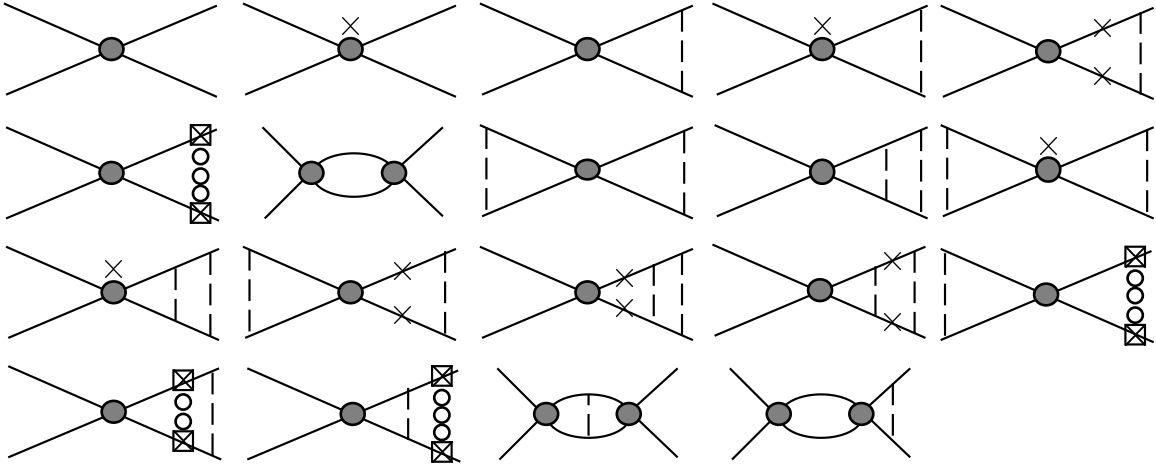}}
\caption{ 
NRQED diagrams for the matching calculation. Permutations are not
displayed.
}
\label{fig2}
\end{figure}

For the NRQED calculation we start from the NRQED
Lagrangian~\cite{Caswell1}
\begin{eqnarray}
\lefteqn{
{\cal{L}}_{\mbox{\tiny NRQED}} \, = \,
-\frac{1}{2}\,(\,{\mbox{\boldmath $E$}}^2-
    {\mbox{\boldmath $B$}}^2\,)
+ \psi^\dagger\,\bigg[\,
i D_t 
+ \frac{{\mbox{\boldmath $D$}}^2}{2\,m_e} 
+ c_1\,\frac{{\mbox{\boldmath $D$}}^4}{8\,m_e^3}
\,\bigg]\,\psi + \ldots
} \nonumber\\[2mm] & &
+ \,\psi^\dagger\,\bigg[\,
\frac{c_1\,e}{2\,m_e}\,{\mbox{\boldmath $\sigma$}}\cdot
    {\mbox{\boldmath $B$}}
+ \frac{c_3\,e}{8\,m_e^2}\,(\,{\mbox{\boldmath $D$}}\cdot 
  {\mbox{\boldmath $E$}}-{\mbox{\boldmath $E$}}\cdot 
  {\mbox{\boldmath $D$}}\,)
+ \frac{c_4\,e}{8\,m_e^2}\,i\,(\,{\mbox{\boldmath $D$}}\times 
  {\mbox{\boldmath $E$}}-{\mbox{\boldmath $E$}}\times 
  {\mbox{\boldmath $D$}}\,)
\,\bigg]\,\psi 
+ \ldots \nonumber\\[2mm] & &
- \,\frac{d_1\,e^2}{4\,m_e^2}\,
  (\psi^\dagger{\mbox{\boldmath $\sigma$}}\sigma_2\chi^*)\,
  (\chi^T\sigma_2{\mbox{\boldmath $\sigma$}}\psi)
+ \frac{d_2\,e^2}{3\,m_e^4}\,
  \frac{1}{2}\Big[\,
  (\psi^\dagger{\mbox{\boldmath $\sigma$}}\sigma_2\chi^*)\,
  (\chi^T\sigma_2{\mbox{\boldmath $\sigma$}}
    (-\mbox{$\frac{i}{2}$}
  {\stackrel{\leftrightarrow}{\mbox{\boldmath $D$}}})^2\psi)
  +\mbox{H.c.}\,\Big]
+\ldots\,,
\label{Lagrangian}
\end{eqnarray}
where $\psi$ and $\chi$ are the electron and positron Pauli spinors
and $D_t$ and ${\mbox{\boldmath $D$}}$ the time and space components of the
gauge covariant derivative $D_\mu$. In Eq.~(\ref{Lagrangian}) the
straightforward bilinear positron terms are omitted and only those
four fermion interaction relevant for the one photon annihilation
contributions to the hfs
are displayed. The renormalization constants $c_1,\ldots,c_4,d_1,d_2$
are normalized to one at the Born level. For $W^{\mbox{\tiny
1-$\gamma$ ann}}_{\mbox{\tiny NNLO}}$ only the radiative corrections to $d_1$
have to be calculated. 
To order $m_e \alpha^6$, and if we consider only the contributions
from the one photon annihilation graphs, 
all retardation effects can be neglected. This means that  
the transverse photon propagators can be used in the
instantaneous approximation, {\it i.e.} their energy dependence is
dropped. Indeed, simple counting rules~\cite{MQED} show that retardation
corrections to the one photon annihilation diagrams would set in at 
order $m_e \alpha^7$.  In the instantaneous approximation, all NRQED
interactions can be written as  two-body potentials. 
In momentum space representation all the potentials needed 
for the present calculation are given by (see Fig.~\ref{fig1})  
\begin{eqnarray}
V_c({\mbox{\boldmath $p$}},{\mbox{\boldmath $q$}}) & = & 
-\,\frac{e^2}{|{\mbox{\boldmath $p$}}-
{\mbox{\boldmath $q$}}|^2+\lambda^2}\,,
\qquad \mbox{(Coulomb interaction)}
\\[2mm]
V_{\mbox{\tiny rel}}({\mbox{\boldmath $p$}},
  {\mbox{\boldmath $q$}}) & = & 
-\,\frac{e^2}{m_e^2}\,\bigg[\,
\frac{|{\mbox{\boldmath $p$}\times \mbox{\boldmath $q$}}|^2} 
   {(|{\mbox{\boldmath $p$}}-{\mbox{\boldmath $q$}}|^2+\lambda^2)^2} 
- \frac{({\mbox{\boldmath $p$}}-{\mbox{\boldmath $q$}})\times 
   {\mbox{\boldmath $S_{-}$}}\cdot 
   ({\mbox{\boldmath $p$}}-{\mbox{\boldmath $q$}})\times 
   {\mbox{\boldmath $S_{+}$}}}
   {|{\mbox{\boldmath $p$}}-{\mbox{\boldmath $q$}}|^2+\lambda^2}
\nonumber\\[2mm] & & \qquad\quad
+ \,i\,\frac{3}{2}\,\frac{({\mbox{\boldmath $p$}}\times 
   {\mbox{\boldmath $q$}})\cdot({\mbox{\boldmath $S_{-}$}}+
   {\mbox{\boldmath $S_{+}$}})}
   {|{\mbox{\boldmath $p$}}-{\mbox{\boldmath $q$}}|^2+\lambda^2}
- \frac{1}{4}\,\frac{|{\mbox{\boldmath $p$}}-
   {\mbox{\boldmath $q$}}|^2}
   {|{\mbox{\boldmath $p$}}-{\mbox{\boldmath $q$}}|^2+\lambda^2}
\,\bigg]
\,,
\\[2mm]
V_4({\mbox{\boldmath $p$}},{\mbox{\boldmath $q$}}) & = & 
\frac{e^2}{2 m_e^2}\,\bigg[\,
\frac{3}{4} + {\mbox{\boldmath $S_{-}$}}\cdot 
  {\mbox{\boldmath $S_{+}$}}
\,\bigg]
\,,
\\[2mm]
V_{4\mbox{\tiny der}}({\mbox{\boldmath $p$}},
  {\mbox{\boldmath $q$}}) & = &  
-\,\frac{e^2}{3 m_e^4}\,({\mbox{\boldmath $p$}}^2+
  {\mbox{\boldmath $q$}}^2)\,\bigg[\,
\frac{3}{4} + {\mbox{\boldmath $S_{-}$}}\cdot 
  {\mbox{\boldmath $S_{+}$}}
\,\bigg]
\,,
\end{eqnarray}
where $\lambda$ is a small fictitious photon mass to regularize IR
divergences and ${\mbox{\boldmath $S_{\mp}$}}$ denotes the 
electron/positron spin
operator. $V_{\mbox{\tiny rel}}$ denotes the $1/m_e^2$
corrections to the Coulomb potential including longitudinal and
transverse photon exchange.
$V_4$ accounts for the leading order
annihilation process $e^+e^-\to\gamma\to e^+e^-$ and $V_{4\mbox{\tiny
der}}$ denotes relativistic corrections to $V_4$ from the energy
dependence of the annihilation photon and from the $1/m_e^2$
contributions in the Dirac spinors.
The calculation of
 $W^{\mbox{\tiny 1-$\gamma$ ann}}_{\mbox{\tiny NNLO}}$ proceeds
in two basic steps. 
\begin{itemize}
\item[1.]{\it Matching calculation:} calculation of the
${\cal{O}}(\alpha)$ and ${\cal{O}}(\alpha^2)$ contributions to the
constant $d_1$ by matching the NRQED and QED amplitudes for the
elastic s-channel scattering of free and on-shell electrons and
positrons via a single photon, $e^+e^-\to\gamma\to e^+e^-$, up to two
loops and to NNLO in the velocity of the electrons and positrons in
the c.m.~frame. 
\item[2.] {\it Bound state calculation:} calculation of
$W^{\mbox{\tiny 1-$\gamma$ ann}}_{\mbox{\tiny NNLO}}$ by solving the
non-relativistic bound state problem in form of the Schr\"odinger
equation (i.e. including the non-relativistic kinetic energy and the
Coulomb interaction) exactly and by treating the relativistic effects
using Rayleigh-Schr\"odinger time-independent perturbation theory (TIPT).
\end{itemize} 
{\it Matching calculation:} to determine the NRQED amplitude
$e^+e^-\to\gamma\to e^+e^-$ for free and on-shell electrons and
positrons up to two loops and NNLO in the velocity the diagrams
displayed in Fig.~\ref{fig2} have to be calculated. 
It is sufficient
to consider only scattering of the $e^+e^-$ pair in a ${}^3\!S_1$
state because a ${}^1\!S_0$ state cannot annihilate into a single photon
due to $C$ invariance. We have regularized all UV divergent
integrations by using a momentum cut-off. As a consequence the finite
terms in the NRQED amplitude (and also in the constant $d_1$) depend
on the routing of the loop momenta through the diagrams~\cite{thesis}.
Therefore, to
be consistent exactly the same way of routing has to be used in the
bound state calculation. We come back to this point later. The
corresponding QED amplitude for the scattering process has to be
determined by using conventional covariant multi-loop perturbation
theory. Whereas the one-loop results for the vertex
corrections~\cite{Schwinger1} and the one and two-loop contributions
to the vacuum polarization function~\cite{Kallensabry1} have been
known for quite a long 
time, the two-loop vertex corrections have been calculated recently by
one of the authors~\cite{Hoang1}. The QED amplitude is renormalized by the
common on-shell renormalization scheme, where $\alpha$ is the fine
structure constant and the wave function renormalization constant is
fixed by the requirement that the residue of the fermion propagator is
one. The ${\cal{O}}(\alpha)$ and ${\cal{O}}(\alpha^2)$ contributions
of the renormalization constant $d_1$ are then determined by demanding
equality of the NRQED and QED amplitude at the one- and the two-loop
level. Because all IR divergent and velocity dependent contributions
are equal in NRQED and QED, $d_1$ contains only UV divergent and
constant contributions. \\
{\it Bound state calculation:} to finally determine $W^{\mbox{\tiny
1-$\gamma$ ann}}_{\mbox{\tiny NNLO}}$ we start from the well known
solution of 
the non-relativistic positronium problem (in form of the Schr\"odinger
equation) and incorporate $V_4$, $V_{4\mbox{\tiny
der}}$, $V_{\mbox{\tiny rel}}$ and $V_{\mbox{\tiny kin}}$ via
first and second order TIPT. For each insertion of $V_4$, the
contributions from $d_1$ have also to be taken into account. The
divergences in $d_1$ automatically remove the UV divergences which
arise in the bound state calculation. At this point we want to
emphasize again that, to be consistent, the routing of the momenta in
the bound state calculation has to be exactly the same as the routing
in the NRQED scattering diagrams. Also the finite terms in the bound
state integrals depend on the routing. Combining the result of the
bound state integrals with the contributions in $d_1$ leads to the
cancellation of the routing-dependent terms. We have checked this fact
by choosing different routings in our calculation. \\
The final result for the order $m_e \alpha^6$ one photon annihilation
contributions to the hfs reads\footnote{
The contributions in
 $W^{\mbox{\tiny 1-$\gamma$ ann}}_{\mbox{\tiny NNLO}}$ coming
from the vacuum polarization effects of the annihilation photon have
been calculated before in~\cite{Barbieri1, Hoang2}. The result
in~\cite{Barbieri1}
contains an error in the treatment of the one-loop vacuum polarization
(see~\cite{Hoang2}). The vacuum polarization contributions calculated
in~\cite{Hoang2} are in agreement with our result.
}
\begin{equation}
W^{\mbox{\tiny 1-$\gamma$ ann}}_{\mbox{\tiny NNLO}} \, = \,
\frac{m_e \alpha^6}{4}\,\bigg[\,
\frac{1}{\pi^2}\,\bigg(\,
\frac{1477}{81} + \frac{13}{8}\,\zeta_3
\,\bigg)
- \frac{1183}{288} + \frac{9}{4}\,\ln 2 + 
\frac{1}{6}\,\ln\alpha^{-1}
\,\bigg]
\,.
\label{final}
\end{equation}
The $\ln\alpha^{-1}$ term was already known and is included in the $\ln\alpha^{-1}$
contribution quoted in Eq.~(\ref{LOandNLO}). 
The one photon annihilation contribution to the constant $K$
corresponds to a contribution of $-2.34$~Mhz to the theoretical
prediction of the hfs (see Table~\ref{tab1}).
\par
The second, more intuitive  method to determine $W^{\mbox{\tiny
1-$\gamma$ ann}}_{\mbox{\tiny NNLO}}$ starts from the formal result for the
energy shift due to one photon annihilation  for S-wave 
triplet bound states with radial quantum numbers $n$ using first to
third order TIPT\footnote{ 
${}^1\!S_0$ states do not contribute to
 $W^{\mbox{\tiny 1-$\gamma$ ann}}_n$ because
they cannot annihilate into a single photon. States with higher
orbital angular momentum, on the other hand, are irrelevant because
their wave functions vanish for zero electron-positron distance.
},
\begin{eqnarray}
W^{\mbox{\tiny 1-$\gamma$ ann}}_{n} & = &
\langle\,n\,| \, V_4 + V_{4\mbox{\tiny der}}  
+ V_4 \, \sum\hspace{-5.5mm}\int\limits_{l\ne n}\,
\frac{|\,l\,\rangle\,\langle \,l\,|}{E_n-E_l} \, V_4 
+ V_4 \, \sum\hspace{-6mm}\int\limits_{k\ne n}
\frac{|\,k\,\rangle\,\langle \,k\,|}{E_n-E_k} \, V_4 \,
 \sum\hspace{-6.5mm}\int\limits_{m\ne n}
\frac{|\,m\,\rangle\,\langle \,m\,|}{E_n-E_m} \, V_4 
\, |\,n\,\rangle  
+\ldots
\nonumber\\[2mm] & &
+\,\bigg[\,
\langle\,n\,| \,
V_4 \, \sum\hspace{-5.5mm}\int\limits_{l\ne n}\,
\frac{|\,l\,\rangle\,\langle \,l\,|}{E_n-E_l} \,
(V_{\mbox{\tiny rel}}+V_{\mbox{\tiny kin}})
\, |\,n\,\rangle + \mbox{H.c.}
\,\bigg]
\nonumber\\[2mm] & &
+\,\bigg[\,
\langle\,n\,| \,
V_4 \, \sum\hspace{-6mm}\int\limits_{k\ne n}\,
\frac{|\,k\,\rangle\,\langle \,k\,|}{E_n-E_k} \,
V_4 \, \sum\hspace{-6.5mm}\int\limits_{m\ne n}\,
\frac{|\,m\,\rangle\,\langle \,m\,|}{E_n-E_m} \,
(V_{\mbox{\tiny rel}}+V_{\mbox{\tiny kin}})
\, |\,n\,\rangle + \mbox{H.c.}
\,\bigg]
+\ldots
\,,
\label{formal}
\end{eqnarray} 
where $|\,i\,\rangle$, $i=l,m,k$, represent normalized (bound state
and continuum) eigenfunctions to the positronium Schr\"odinger
equation with the eigenvalues $E_i$. It is evident from the form
of the operator $V_4$ that $W^{\mbox{\tiny 1-$\gamma$ ann}}_{n}$
depends only on the zero-distance Coulomb Green function 
$
A_n\equiv \langle\,\vec{0}\,| \,
\sum_{l\ne n}\hspace{-0.9cm}\int\hspace{0.5cm}\,
\frac{|\,l\,\rangle\,\langle \,l\,|}{E_l-E_n}
\, |\,\vec{0}\,\rangle
$
(where the $n{}^3\!S_1$ bound state pole is
subtracted) and on the rate for annihilation of a $n{}^3\!S_1$ bound
state into a single photon,
$
P_n\equiv
\langle\,n\,| \, V_4 \, |\,n\,\rangle +
[
\langle\,n\,| \, V_4 \,\sum_{l\ne n}\hspace{-0.9cm}\int\hspace{0.5cm}\,
\frac{|\,l\,\rangle\,\langle \,l\,|}{E_n-E_l} \, 
(V_{\mbox{\tiny rel}}+V_{\mbox{\tiny kin}})
\,|\,n\,\rangle + \mbox{H.c.}
] +
\langle\,n\,| \, V_{4\mbox{\tiny der}}
\, |\,n\,\rangle 
$
(where the effects from $V_{\mbox{\tiny rel}}$ and $V_{\mbox{\tiny
kin}}$ are included in form of corrections
to the wave function). Because $A_n$ and $P_n$ are UV divergent from the
integration over the high energy modes, they have to be
renormalized. In the NRQED approach this was achieved by the
renormalization constant $d_1$. Here, renormalization will be carried
out by relating $A_n$ and $P_n$ to physical (and finite) quantities which
incorporate the proper short-distance physics from the one photon
annihilation
process. For $A_n$ this physical quantity is just the QED vacuum
polarization function in the non-relativistic limit and for $P_n$ the
abelian contribution of the NNLO expression for the leptonic decay
width of a superheavy quark-antiquark $n{}^3\!S_1$ bound
state~\cite{Bodwin1}. Both 
quantities have been determined recently in~\cite{Hoang2,
Hoang3}. From the results of~\cite{Hoang2, Hoang3} it is
straightforward to derive the 
renormalized versions of $A_n$ and $P_n$,
\begin{eqnarray}
A^{\mbox{\tiny phys}}_n & = &
\frac{m_e^2}{2\,\pi}\,\bigg\{\,
\frac{8}{9\,\pi} - \frac{\alpha}{2}\,\bigg[\,
C_1 + 
\bigg(\, \ln\Big(\frac{\alpha}{2\,n}\Big) - \frac{1}{n} + 
\gamma + \Psi(n)
\,\bigg)
\,\bigg]
\,\bigg\}
\,,
\\[2mm]
P^{\mbox{\tiny phys}}_n & = &
\frac{2\,\alpha\,\pi}{m_e^2}\,
\bigg(\frac{m_e^3\,\alpha^3}{8\,\pi\,n^3}\bigg)\,
\bigg\{\,
1 - 4\,\frac{\alpha}{\pi} + 
\alpha^2\,\bigg[\, C_2 - \frac{37}{24\,n^2} - 
\frac{2}{3}\,\bigg(\, \ln\Big(\frac{\alpha}{2\,n}\Big) - \frac{1}{n} + 
\gamma + \Psi(n)
\,\bigg)
\,\bigg]
\,\bigg\}
\,,
\label{renormalized}
\end{eqnarray}
where 
$
C_1  = 
\frac{1}{2\pi^2}(-3+\frac{21}{2}\zeta_3
) - \frac{11}{16} + \frac{3}{2}\ln 2
$
and
$
C_2  =  
\frac{1}{\pi^2}(\frac{527}{36}-\zeta_3) + 
\frac{4}{3}\ln 2 - \frac{43}{18}
$.
$\gamma$ is the Euler constant and $\Psi$ the digamma
function.
Inserting $A^{\mbox{\tiny phys}}_n$ and $P^{\mbox{\tiny phys}}_n$ back
into expression~(\ref{formal}) we arrive at
\begin{equation}
W^{\mbox{\tiny 1-$\gamma$ ann}}_{n} \, = \, 
P^{\mbox{\tiny phys}}_n\,\bigg[\,
1- \frac{2\,\alpha\,\pi}{m_e^2}\,A^{\mbox{\tiny phys}}_n +
\bigg(\,\frac{2\,\alpha\,\pi}{m_e^2}\,A^{\mbox{\tiny phys}}_n\,\bigg)^2 
\,\bigg]
\,.
\label{finalsecond}
\end{equation}
It is an easy task to check for the ground state $n=1$ that in
Eq.~(\ref{finalsecond}) the well known
order $m_e\alpha^4$ ($W^{\mbox{\tiny 1-$\gamma$ ann}}_{\mbox{\tiny
LO}}=\frac{1}{4}\,m_e\alpha^4$) and 
$m_e\alpha^5$ ($W^{\mbox{\tiny 1-$\gamma$ ann}}_{\mbox{\tiny
NLO}}=-\frac{11}{9\pi}\,m_e\alpha^5$)  one photon annihilation 
contributions to the hfs are correctly reproduced and that the
$m_e\alpha^6$ contributions are equal to $W^{\mbox{\tiny
1-$\gamma$ ann}}_{\mbox{\tiny NNLO}}$, Eq.~(\ref{final}).
This second, very simple intuitive method to determine 
$W^{\mbox{\tiny 1-$\gamma$ ann}}_{\mbox{\tiny NNLO}}$
does not only represent a nice cross check
for the systematic NRQED calculation but also illustrates that 
$W^{\mbox{\tiny 1-$\gamma$ ann}}$ is directly related to
other physical quantities. Beyond order $m_e\alpha^6$
expression~(\ref{finalsecond}) is not valid because essential
retardation effects are not taken into account.
\par
\begin{table}[h]
\begin{center}
\begin{tabular}{rllcrl} \hline
&       &               & analytical/ &  & \\
& \raisebox{1.5ex}[-1.5ex]{Order} 
& \raisebox{1.5ex}[-1.5ex]{Specification} & numerical   
& \raisebox{1.5ex}[-1.5ex]{Contr. in Mhz} 
& \raisebox{1.5ex}[-1.5ex]{Refs.} 
\\ \hline
1. & $m_e \alpha^4$ & & a & $204\,386.7(1)\mbox{\hspace{2mm}}$ & \cite{Pirenne1} 
\\
2. & $m_e \alpha^5$ & & a & $-1\,005.5\mbox{\hspace{7mm}}$ & \cite{Karplus1}
\\
3. & $m_e \alpha^6 \ln\alpha^{-1}$ & & a & $19.1\mbox{\hspace{7mm}}$ &
    \cite{CaswellBodwin} 
\\
4. & $m_e \alpha^6$ & non-annihilation (C/L) & n & 
    $-7.2(6)\mbox{\hspace{2mm}}$ &
    \cite{Sapirstein1}+\cite{onephoton}+\cite{Caswell1} 
\\
5. &  & non-annihilation (Pa) & n & $-3.29(4)$ &
    \cite{Sapirstein1}+\cite{onephoton}+\cite{Pachucki1}
\\
6. &  & 1 photon annihilation & a & $-2.34\mbox{\hspace{5mm}}$ &
    this work 
\\
7. &  & 2 photon annihilation & a & $-0.61\mbox{\hspace{5mm}}$ &
    \cite{Adkins1}
\\
8. &  & 3 photon annihilation & a & $-0.97\mbox{\hspace{5mm}}$ &
    \cite{Adkins2}
\\
9. & $m_e \alpha^7 \ln^2\alpha^{-1}$ &  & a & $-0.92\mbox{\hspace{5mm}}$ &
    \cite{Karshenboim1,thesis}
\\ \hline
& & Sum (Caswell-Lepage) & & $203\,388.3(6)\mbox{\hspace{2mm}}$ &
\\ 
& & Sum (Pachucki)  & & $203\,392.2(1)\mbox{\hspace{2mm}}$ &
\\ 
& & Experiment  & & $203\,389.1(7)\mbox{\hspace{2mm}}$ &
  \cite{Ritter1}
\\ \hline
\end{tabular}
\parbox{15.6cm}{
\caption{\label{tab1}
Summary of the theoretical calculations to the hfs. Only the
references with the first correct calculations are given.} 
}
\end{center}
\end{table}
In Table~\ref{tab1} we have summarized the status of the theoretical
calculation to the hfs of the positronium ground state including our
own result. To order $m_e \alpha^6$ the logarithmic in $\alpha$ and
constant contributions are given separately. The constant terms are
further subdivided into non-annihilation, and one, two and
three photon annihilation contributions. The error in the order $m_e
\alpha^4$ result (1.) comes from the uncertainties in the input
parameters $\alpha$, $\hbar$ and $m_e$ and the errors in 4. and 5. are
numerical. For all other contributions the errors are negligible. As
indicated, there are two contradictory calculations for some of 
non-annihilation contributions based on results from Caswell and
Lepage (C/L)~\cite{Caswell1} and Pachucki (Pa)~\cite{Pachucki1}. 
The result containing the Caswell-Lepage calculation leads to
perfect agreement between theory and experiment ($W_{\mbox{\tiny
th}}-W_{\mbox{\tiny ex}}=-0.8(1.0)$~Mhz), whereas the hfs prediction
based on the result by Pachucki
leads to a discrepancy of more than four standard
deviations ($W_{\mbox{\tiny th}}-W_{\mbox{\tiny ex}}=3.1(0.7)$~Mhz). 
It remains the task of future examinations to finally
resolve the theoretical situation.  
\par
During completion of this work we were informed of work on the same 
subject by Adkins, Fell and Mitrikov using the Bethe-Salpeter
formalism and numerical methods. Their result agrees with ours
representing an independent cross check. We thank G. Adkins and his
group for reporting their result to us prior to publication. 
We also thank G.P.~Lepage for useful discussions.
This work is supported in part by
the Department of Energy under contract DOE~DE-FG03-90ER40546
and by the Natural Sciences and Engineering Research Council 
of Canada.

%
%
\sloppy
\raggedright
\def\app#1#2#3{{\it Act. Phys. Pol. }{\bf B #1} (#2) #3}
\def\apa#1#2#3{{\it Act. Phys. Austr.}{\bf #1} (#2) #3}
\def\lhc{Proc. LHC Workshop, CERN 90-10}
\def\npb#1#2#3{{\it Nucl. Phys. }{\bf B #1} (#2) #3}
\def\nP#1#2#3{{\it Nucl. Phys. }{\bf #1} (#2) #3}
\def\plb#1#2#3{{\it Phys. Lett. }{\bf B #1} (#2) #3}
\def\prd#1#2#3{{\it Phys. Rev. }{\bf D #1} (#2) #3}
\def\pra#1#2#3{{\it Phys. Rev. }{\bf A #1} (#2) #3}
\def\pR#1#2#3{{\it Phys. Rev. }{\bf #1} (#2) #3}
\def\prl#1#2#3{{\it Phys. Rev. Lett. }{\bf #1} (#2) #3}
\def\prc#1#2#3{{\it Phys. Reports }{\bf #1} (#2) #3}
\def\cpc#1#2#3{{\it Comp. Phys. Commun. }{\bf #1} (#2) #3}
\def\nim#1#2#3{{\it Nucl. Inst. Meth. }{\bf #1} (#2) #3}
\def\pr#1#2#3{{\it Phys. Reports }{\bf #1} (#2) #3}
\def\sovnp#1#2#3{{\it Sov. J. Nucl. Phys. }{\bf #1} (#2) #3}
\def\sovpJ#1#2#3{{\it Sov. Phys. JETP Lett. }{\bf #1} (#2) #3}
\def\jl#1#2#3{{\it JETP Lett. }{\bf #1} (#2) #3}
\def\jet#1#2#3{{\it JETP }{\bf #1} (#2) #3}
\def\zpc#1#2#3{{\it Z. Phys. }{\bf C #1} (#2) #3}
\def\ptp#1#2#3{{\it Prog.~Theor.~Phys.~}{\bf #1} (#2) #3}
\def\nca#1#2#3{{\it Nuovo~Cim.~}{\bf #1A} (#2) #3}
\def\ap#1#2#3{{\it Ann. Phys. }{\bf #1} (#2) #3}
\def\hpa#1#2#3{{\it Helv. Phys. Acta }{\bf #1} (#2) #3}
\def\ijmpA#1#2#3{{\it Int. J. Mod. Phys. }{\bf A #1} (#2) #3}
\def\ZETF#1#2#3{{\it Zh. Eksp. Teor. Fiz. }{\bf #1} (#2) #3}
\def\jmp#1#2#3{{\it J. Math. Phys. }{\bf #1} (#2) #3}
\def\yf#1#2#3{{\it Yad. Fiz. }{\bf #1} (#2) #3}

%
\end{document}